\documentclass[aps,pre,twocolumn,nofootinbib,floatfix]{revtex4}
\usepackage{amsmath}
\usepackage{epsfig}

\input{tcilatex}

\begin{document}

\title{Renormalized frequency shift of a Wannier exciton in a
one-dimensional system}
\author{Yueh-Nan Chen}
\email{ynchen.ep87g@nctu.edu.tw}
\author{Der-San Chuu}
\email{dschuu@cc.nctu.edu.tw }
\affiliation{Department of Electrophysics, National Chiao-Tung University, Hsinchu 300,
Taiwan}
\date{\today }

\begin{abstract}
The radiative frequency shift of superradiant exciton in a one-dimensional
system is calculated. It is shown that\ a finite frequency shift can be
obtained after proper renormalization. The value of the shift is inversely
proportional to the factor $\frac{\lambda }{d}$ , where $\lambda $ is the
wavelength of the emitted photon and $d$ is the lattice spacing.

\end{abstract}

\maketitle

\address{Department of Electrophysics, National Chiao Tung University,
Hsinchu 30050, Taiwan}

\address{Department of Electrophysics, National Chiao Tung University,
Hsinchu 30050, Taiwan}

\address{Department of Electrophysics, National Chiao Tung University,
Hsinchu 300, Taiwan}

%\draft
%\wideabs{

% version ZI

%}

%%%%%%%%%%%%%%%%%%%%%%%%%%%%%%%%%%%%%%%%%%%%%%%%%%%%%%%%%%

Since Dicke proposed the phenomenon of superradiance\cite{1}, the coherent
effect for spontaneous radiation of various systems has attracted extensive
interest both theoretically and experimentally\cite{2,3}. In semiconductor
systems, the electron-hole pair is naturally a candidate for examining the
spontaneous emission. However, as it was well known, the excitons in a three
dimensional system will couple with photons to form polaritons--the
eigenstate of the combined system consisting of the crystal and the
radiation field which does not decay radiatively\cite{4}. Thus, in a bulk
crystal, the exciton can only decay via impurity, phonon scatterings, or
boundary effects.

The exciton can render radiative decay in lower dimensional systems such as
quantum wells, quantum wires, or quantum dots as a result of broken
symmetry. In 1966, V. M. Agranovich \emph{et al.} predicted that the decay
rate of the exciton is superradiantly enhanced by a factor of $(\lambda
/d)^{2}$ for a 2D exciton-polariton system\cite{5}, where $\lambda $ is the
wave length of the emitted photon and $d$ is the lattice constant of the
thin film. First observation of superradiant short lifetimes of excitons was
performed by Y. Aaviksoo \emph{et al.} on surface states of the anthracene%
\cite{6}. In the past decades, the superradiance of excitons in these
quantum well structures has been investigated intensively\cite{7,8,9,10,11}.
For lower dimensional systems, the decay rate of the exciton is enhanced by
a factor of $\lambda /d$ in a quantum wire\cite{5}. In the quantum dots, the
decay rate is shown to be proportional to R$^{2.1}$\cite{12} which confirms
the theoretical prediction\cite{13,14}. In fact, superradiance is
accompanied by frequency shift, as pointed out in Ref. \cite{15}. Although
the spectrum of polaritons in one-dimensional state was studied in Refs.\cite%
{5,16}, the radiative correction to the frequency of the spontaneous
radiation from the Wannier exciton in a one-dimensional system has not been
displayed explicitly. The reason is that the radiation correction to the
frequency shift usually contains divergences which have to be removed by
renormalization. In this paper, we show that the radiative frequency shift
of the Wannier exciton can be properly renormalized in a one-dimensional
system. The renormalized frequency shift of the Wannier exciton in 1-D is
also superradiatively enhanced due to the coherent effect.

Consider now a Wannier exciton in a one-dimensional system with lattice
spacing $d$. We will assume a two-band model for the band structure of the
system. The state of the Wannier exciton with coherent length $L_{c}$ can be
phenomenologically approximated as

\begin{equation}
\left| k_{z},n\right\rangle =\sum_{l\rho }\sqrt{\frac{d}{L_{c}}}\exp
(ik_{z}r_{c})F_{n}(l),
\end{equation}
where $k_{z}$ is the crystal momentum on the chain direction characterizing
the motion of the exciton, $n$ is the quantum number for the internal
structure of the exciton, and, in the effective mass approximation, $r_{c}=%
\frac{m_{e}^{\ast }(l+\rho )+m_{h}^{\ast }\rho }{m_{e}^{\ast }+m_{h}^{\ast }}
$ is the center of mass of the exciton. $F_{n}(l)$ is the hydrogenic wave
function with $l+\rho $ and $\rho $ being the positions of the electron and
hole, respectively. Here $m_{e}^{\ast }$ and $m_{h}^{\ast }$ are,
respectively, the effective masses of the electron and hole. If one neglects
the effects of imperfections and scatterings, the coherent length in Eq. (1)
should become infinite.

The interaction between the exciton and the photon can be written in the form

\begin{equation}
H^{\prime }=\sum_{k_{z}n}\sum_{\mathbf{q}^{\prime }k_{z}^{\prime }}D_{%
\mathbf{q}^{\prime }k_{z}^{\prime }k_{z}n}b_{k_{z}^{\prime }\mathbf{q}%
^{\prime }}c_{k_{z}n}^{\dagger }+\mathbf{H.c.,}
\end{equation}
where

\begin{equation}
D_{\mathbf{q}^{\prime }k_{z}^{\prime }k_{z}n}=\frac{e}{mc}\sqrt{\frac{2\pi
\hbar cL_{c}}{(q^{\prime 2}+k_{z}^{\prime 2})^{1/2}dv}}\epsilon _{\mathbf{q}%
^{\prime }k_{z}^{\prime }}\chi _{k_{z}n}
\end{equation}
with $\epsilon _{\mathbf{q}^{\prime }k_{z}}$ being the polarization of the
photon. $c_{k_{z}n}$ and $b_{\mathbf{q}^{\prime }k_{z}^{\prime }}$ are the
operators of the exciton and photon, respectively. $E_{k_{z}n}$ is the
exciton dispersion. In Eq. (3),

\begin{eqnarray}
\chi _{k_{z}n} &=&\sum_{l}F^{\ast }(l)\int d\tau \omega _{c}(\tau -l)\exp
(ik_{z}(\tau -\frac{m_{e}^{\ast }}{m_{e}^{\ast }+m_{h}^{\ast }})l)  \notag \\
&&\times (-i\hbar \frac{\partial }{\partial \tau })\omega _{v}(\tau )
\end{eqnarray}%
is the effective transition dipole matrix element between the electronic
Wannier state $\omega _{c}$ in the conduction band and the Wannier hole
state $\omega _{v}$ in the valence band.

Because of the presence of exciton-photon interaction, the radiative decay
of the exciton is expected to keep dual merits of coherent nature in the
wire direction and superradiant decay perpendicular to the wire. In the
interaction picture, the state $\left| \psi (t)\right\rangle $ for the whole
system composed of the exciton (with frequency shift $\Omega _{k_{z}n}$ and
finite decay rate $\gamma _{_{k_{z}n}}$ ) and photons can be written as

\begin{equation}
\left| \psi (t)\right\rangle =e^{-i\Omega _{k_{z}n}t-\frac{1}{2}\gamma
_{_{k_{z}n}}t}\left| k_{z},n;0\right\rangle +\sum_{\mathbf{q}^{\prime
}k_{z}^{\prime }}f_{G;\mathbf{q}^{\prime }k_{z}^{\prime }}(t)\left| G;%
\mathbf{q}^{\prime }k_{z}^{\prime }\right\rangle ,
\end{equation}
where $\left| k_{z},n;0\right\rangle $ is the state with a Wannier exciton
in the mode $k_{z},n$ in the linear chain without photons, and $\left| G;%
\mathbf{q}^{\prime }k_{z}^{\prime }\right\rangle $ represents the state in
which the electron-hole pair recombines and a photon in the mode $\mathbf{q}%
^{\prime }\mathbf{,}k_{z}^{\prime }$ is created.

The decay rate and frequency shift can be evaluated as\cite{17}

\begin{equation}
\gamma _{_{k_{z}n}}=2\pi \sum_{\mathbf{q}^{\prime }k_{z}^{\prime }}\left| D_{%
\mathbf{q}^{\prime }k_{z}^{\prime }k_{z}n}\right| ^{2}\delta (\omega _{%
\mathbf{q}^{\prime }k_{z}^{\prime }k_{z}n})
\end{equation}
and

\begin{equation}
\Omega _{k_{z}n}=\mathcal{P}\sum_{\mathbf{q}^{\prime }k_{z}^{\prime }}\frac{%
\left| D_{\mathbf{q}^{\prime }k_{z}^{\prime }k_{z}n}\right| ^{2}}{\omega _{%
\mathbf{q}^{\prime }k_{z}^{\prime }k_{z}n}},
\end{equation}
where $\omega _{\mathbf{q}^{\prime }k_{z}^{\prime }k_{z}n}=E_{k_{z}n}/\hbar
-c\sqrt{q^{\prime 2}+k_{z}^{\prime 2}}$ and $\mathcal{P}$ means the
principal value of the integral. Thus, the Wannier exciton decay rate in the
optical region can be calculated straightforwardly and is given by

\begin{eqnarray}
\gamma _{k_{z}n} &=&\frac{3d}{4k_{0}}\gamma _{0}\int dq^{\prime }\frac{%
\left| \mathbf{\epsilon }_{\mathbf{q}^{\prime }k_{z}^{\prime }}\chi
_{n}\right| }{\left| \chi _{n}\right| ^{2}}^{2}\frac{q^{\prime }}{L_{c}\sqrt{%
k_{0}^{2}-q^{\prime 2}}}  \notag \\
&&\times \frac{\sin ^{2}(L_{c}\sqrt{k_{0}^{2}-q^{\prime 2}}/2)}{\sin ^{2}(d%
\sqrt{k_{0}^{2}-q^{\prime 2}}/2)},
\end{eqnarray}%
where $k_{0}=E_{k_{z}n}/\hbar =2\pi /\lambda $,

\begin{equation}
\chi _{n}=\sum_{l}F_{n}^{*}(l)\int d\tau w_{c}(\tau -l)(-i\hbar \frac{%
\partial }{\partial \tau })w_{v}(\tau ),
\end{equation}
and

\begin{equation}
\gamma _{0}=\frac{4e^{2}\hbar k_{0}}{3m^{2}c^{2}}\left| \chi _{n}\right| ^{2}
\end{equation}%
with $\gamma _{0}$ being the decay rate of an isolated atom. Fig. 1 shows
the numerical calculations of Eq. (8). As can be seen in the figure, the
decay rate of the exciton is increased with the increase of the lateral size
of the exciton center-of-mass wave function, called the exciton coherent
length, and saturates when the coherence length reaches the wavelength $%
\lambda $ of the emitted photon. This is because as $L_{c}>\lambda $, the
superradiant effect becomes prominent, and the decay rate should approach 1D
limit: $\gamma _{1d}=\frac{3\lambda }{2d}\gamma _{0}\frac{\left| \mathbf{%
\epsilon }_{\mathbf{q}^{\prime }k_{z}^{\prime }}\chi _{n}\right| }{\left|
\chi _{n}\right| ^{2}}^{2}.$\cite{5,17} Generally speaking, the enhanced
factor $\lambda /d$ for \emph{perfect} 1D crystal and $(\lambda /d)^{2}$ for
2D film has a simple physical meaning--it is determined by the number of
unit cells in the so-called \emph{cooperative length }$\lambda $. 
\begin{figure}[h]
\includegraphics[width=8cm]{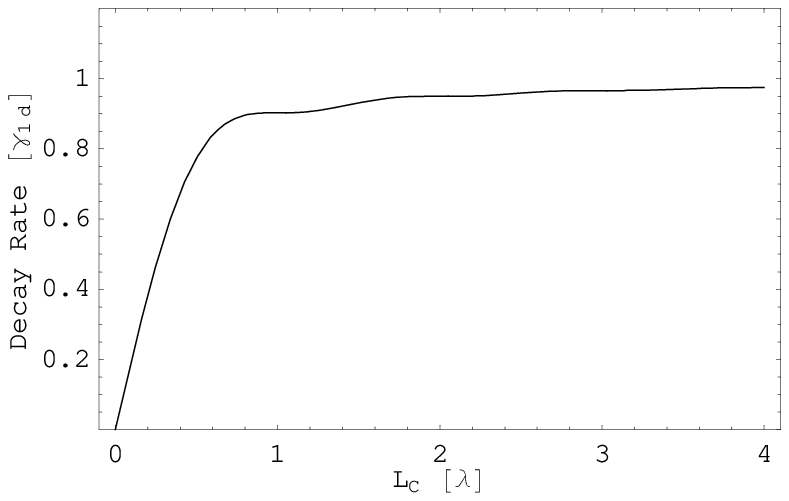}
\caption{Decay rate of an exciton as a function of coherent length $L_{c}$.
The vertical and horizontal units are $\frac{3\protect\lambda }{2d}\protect%
\gamma _{0}\frac{\left| \mathbf{\protect\epsilon }_{\mathbf{q}^{\prime
}k_{z}^{\prime }}\protect\chi _{n}\right| }{\left| \protect\chi _{n}\right|
^{2}}^{2}$ and $\protect\lambda $, respectively. In this graph we have
assumed $\protect\lambda =8000\protect\overset{\circ }{A}$ and lattice
spacing $d=5\protect\overset{\circ }{A}$.}
\end{figure}

Now let us present our results for the renormalized frequency shift. For
simplicity, let us consider the case that the coherent length $L_{c}$ is
much larger than the wavelength of the emitted photon. Thus, the frequency
shift in Eq. (7) can be expressed as

\begin{equation}
\Omega _{k_{z}n}=\frac{2\pi e^{2}\hbar L_{c}}{m^{2}c^{2}dv}\mathcal{P}\sum_{%
\mathbf{q}^{\prime }}\frac{\left| \epsilon _{\mathbf{q}^{\prime }k_{z}}\chi
_{n}\right| ^{2}}{(k_{0}-\sqrt{q^{\prime 2}+k_{z}^{2}})\cdot \sqrt{q^{\prime
2}+k_{z}^{2}}}
\end{equation}%
As can be seen from the above expression, the frequency shift suffers from
infrared divergence when the denominator approaches zero. If one uses the
usual procedure of mass renormalization in atomic physics, it seems the
divergence becomes more strongly. To resolve this question, we adopt the
method of renormalization for a system of two-level atoms proposed by Lee
and Lin\cite{18}. In their works, they identified that the frequency shift
is the radiative level shift and is given by

\begin{equation}
\Delta E_{0}=\mathcal{P}\sum_{\mathbf{q}^{\prime }}\frac{\left\langle
k_{z},n;0\right| H^{\prime }\left| G;\mathbf{q}^{\prime }k_{z}\right\rangle
\left\langle G;\mathbf{q}^{\prime }k_{z}\right| H^{\prime }\left|
k_{z},n;0\right\rangle }{\hbar c(k_{0}-\sqrt{q^{\prime 2}+k_{z}^{2}})},
\end{equation}
where $H^{\prime }$ is defined in Eq. (2). On the other hand, they also
recognized the true radiative correction to the energy level $\alpha $ is

\begin{equation}
\Delta E=\Delta E_{0}+\left\langle \alpha \right| \theta _{c}\left| \alpha
\right\rangle ,
\end{equation}
where $\Delta E_{0}$ is the unrenormalized level shifts, and $\theta _{c}$
is the counter-term operator. Assuming the recoil of the electron upon
emitting or absorbing a photon to be negligible, one can make the following
identification of the operator

\begin{equation}
\theta _{c}=H^{\prime }\frac{1}{H_{ph}}H^{\prime }.
\end{equation}
Therefore, the renormalized result ($k_{z}\sim 0$) can be obtained as

\begin{equation}
\Omega _{k_{z}\sim 0,n}^{ren}=\frac{e^{2}\hbar k_{0}}{m^{2}c^{2}d}\mathcal{P}%
\int_{0}^{k_{m}}\frac{\left| \epsilon _{\mathbf{q}^{\prime }k_{z}}\chi
_{n}\right| ^{2}}{(k_{0}-q^{\prime })q^{\prime }}dq^{\prime },
\end{equation}%
where the upper limit of integration is cut off at $k_{m},$ which is taken
to be the inverse of the electron Compton wavelength as usually done in the
nonrelativistic cases. As one can note from Eq. (15), there is divergent
problem when $q^{\prime }\sim 0$ or $q^{\prime }\sim k_{0}.$ This can be
overcome by substituting $-i\hbar \frac{\partial }{\partial \tau }$ by $%
-imcq^{\prime }\tau $ (Ref. \cite{15}) in Eq. (9) when $q^{\prime }$ is
small. It is equivalent to the dipole-interaction form, $H^{\prime }\sim 
\mathbf{r\cdot E.}$ With this treatment, we have

\begin{equation}
\Omega _{k_{z}\sim 0,n}^{ren}=\mathcal{P}\int_{0}^{k_{m}}B_{q^{\prime
}k_{z}n}dq^{\prime }
\end{equation}
with

\begin{equation}
B_{q^{\prime }k_{z}n}=\left\{ 
\begin{array}{c}
\frac{e^{2}\hbar k_{0}}{m^{2}c^{2}dq^{\prime }}\left| \epsilon _{\mathbf{q}%
^{\prime }k_{z}}\chi _{n}\right| ^{2}\text{, when }q^{\prime }\text{ is large%
} \\ 
\frac{e^{2}\hbar k_{0}q^{\prime }}{d}\left| \epsilon _{\mathbf{q}^{\prime
}k_{z}}\kappa _{n}\right| ^{2}\text{ , when }q^{\prime }\text{ is small}%
\end{array}
\right.
\end{equation}
where

\begin{equation}
\kappa _{n}=\sum_{l}F_{n}^{*}(l)\int d\tau w_{c}(\tau -l)(-i\tau )w_{v}(\tau
).
\end{equation}
In general, the wavelength of the photon is much larger than the Compton
wavelength of the electron. Therefore, the renormalized frequency shift can
be approximated as

\begin{equation}
\Omega _{k_{z}\sim 0,n}^{ren}=-\gamma _{\sin gle}(\frac{1}{k_{0}d}),
\end{equation}
where

\begin{equation}
\gamma _{\sin gle}=\frac{2e^{2}E_{k_{z}\sim 0,n}}{c}\left| k_{0}\kappa
_{n}\right| ^{2}
\end{equation}%
is roughly equal to the radiative decay rate of a single isolated atom. As
can be seen from Eq. (19), the renormalized frequency shift is enhanced by
the factor of $(\frac{1}{k_{0}d})$ in a single quantum wire, while the
frequency shift of the exciton in a thin semiconductor film is inversely
proportional to the square of the factor $k_{0}d$ \cite{15}. We then
conclude that in low dimensional systems the renormalized frequency shift of
the superradiant exciton is enhanced by the factor of $(\frac{1}{k_{0}d}%
)^{x} $, where $x$ is the dimension of the system.

A few remarks about the differences between the previous and our works can
be mentioned here. Since we consider the spontaneous emission of the exciton
in the resonance approximation, the non-resonant interaction between the
exciton and free photons has been omitted. Under this condition, one
immediately meets the divergent problems both in 2D and 1D cases after
performing first order perturbation. In a recent paper by V. V. Popov \emph{%
et al.}\cite{20}, the radiative decay of coherent polariton modes in a
two-dimensional excitonic system is analyzed. The authors state their
approach is suitable for the examinations of time-resolved spontaneous
emission. In fact, this is also the case by using our approach, where the
renormalized procedure is borrowed from atomic physics. For 2D systems and
in the case of normal emission ($k_{\parallel }=0$), the finite
(renormalized) frequency shift from our approach is also superradiant
enhanced and can be written as

\begin{equation}
f_{2D}\approx -\gamma _{0}(\frac{\lambda _{0}}{2\pi d})^{2}L_{z},
\end{equation}%
where $L_{z}$ is the well width, $\lambda _{0}$ is the wavelength of the
emitted photon, $d$ is lattice spacing, and $\gamma _{0}$ is radiative decay
rate of a single isolated exciton. This value agrees well with the V. V.
Popov's calculation in L and T modes\cite{21}

\begin{equation}
\omega -\omega _{0}-i\gamma _{\Gamma }=-\frac{i}{2}k_{z}L_{z}\omega _{LT},
\end{equation}%
where $\omega _{LT}$ means the (superradiant-enhanced) coupling strength.
Since $k_{z}$ is a complex-valued quantity, hence the value $L_{z}\omega
_{LT}($Im$k_{z})/2$ corresponds to our renormalized frequency shift $f_{2D}$%
, where $\gamma _{0}(\frac{\lambda _{0}}{2\pi d})^{2}$ can be viewed as the
(effective) coupling strength between exciton and photons.

For usual semiconductors, the enhanced factor in Eq. (19) is about $10^{3}$
for Wannier excitons in the optical range. However, due to the extreme
smallness of $\gamma _{\sin gle}$ itself, observation of $\Omega _{k_{z}\sim
0,n}^{ren}$ in a single quantum wire is not expected to be easy. If the
decay rate of the exciton is in the order of ps$^{-1}$, the radiative shift
is about $10^{-1}$meV. To observe the coherent effect, one may increase the
number ($N^{\prime }$) of the wires. If $N^{\prime }$ parallel quantum wires
(with $L_{c}>>\lambda $) are placed on the same plane, the state of the
systems is written

\begin{equation}
\left| \psi (t)\right\rangle =\sum_{j=1}^{N^{\prime }}g_{j}(t)\left|
j,k_{z},n;0\right\rangle +\sum_{\mathbf{q}^{\prime }}g_{G;\mathbf{q}^{\prime
}k_{z}}(t)\left| G;\mathbf{q}^{\prime }k_{z}\right\rangle ,
\end{equation}%
where $\left| j,k_{z},n;0\right\rangle $ is the state in which the $j$th
wire is excited with no photon present, and $\left| G;\mathbf{q}^{\prime
}k_{z}\right\rangle $ represents the state in which all wires are unexcited
with one photon present. When the wires are distributed in equal spacing as
in a lattice, the solution can be written as\cite{19}

\begin{equation}
g_{j}(t)=N^{\prime -1}\sum_{p}e^{ipx_{j}}e^{i\varpi (p)t},
\end{equation}
where $x_{j}$ is the position of the $j$th wire, and the $p$ vector is
quantized as usual, obeying the periodic boundary condition. The complex
frequency $\varpi (p)$ can be obtained as

\begin{equation}
\varpi (p)=\sum_{\mathbf{q}^{\prime }}\sum_{j=1}^{N^{\prime }}\left| D_{%
\mathbf{q}^{\prime }k_{z}n}\right| ^{2}\zeta (\omega _{\mathbf{q}^{\prime
}k_{z}n})\exp [i(q_{x}^{\prime }-p)x_{j}],
\end{equation}
where the vector $\mathbf{q}^{\prime }=(q_{x}^{\prime },q_{y}^{\prime })$
and the function $\zeta (\omega _{\mathbf{q}^{\prime }k_{z}n})=\mathcal{P}%
/\omega _{\mathbf{q}^{\prime }k_{z}n}-i\pi \delta (\omega _{\mathbf{q}%
^{\prime }k_{z}n}).$ One can note that from Eq. (23), as $x_{j}\rightarrow 0$%
, the frequency shift $\Omega _{k_{z}n}^{\prime }\equiv \func{Re}\varpi
(p)=N^{\prime }\Omega _{k_{z}n}.$ It means if there are $N^{\prime }$
parallel quantum wires within the \emph{cooperative length }$\lambda
(k_{0}x_{j}=\frac{2\pi }{\lambda }x_{j}<<1)$, the frequency shift will be
enhanced approximately by an extra factor of $N^{\prime }$-- the number of
the wires. Thus, the total enhanced factor to the frequency shift is about $%
N^{\prime }/(k_{0}d)=N^{\prime }\lambda /(2\pi d)$.

In summary, we have studied the radiative decay of the Wannier exciton in a
one-dimensional system. It is shown that the radiative frequency shift can
be explicitly calculated after a proper renormalization has been made.
Similar to its decay-rate counterpart, the renormalized frequency shift is
superradiatively enhanced by the factor of $(\frac{1}{k_{0}d})$.

One of authors (D. S. Chuu) would like to thank to Prof. Y. C. Lee of SUNYAB
for helpfull discussions. This work is supported partially by the National
Science Council, Taiwan under the grant number NSC 91-2120-M-009-002.


\begin{thebibliography}{99}
\bibitem{1} R. H. Dicke, Phys. Rev. \textbf{93}, 99 (1954).

\bibitem{2} N. Skiribanowitz, I. P. Herman, J. C. MacGillivrary, and M. S.
Feld, Phys. Rev. Lett. \textbf{30}, 309 (1973).

\bibitem{3} V. Ernst and P. Stehle, Phys. Rev. 176, \textbf{1456} (1968);F.
Arechi and D. Kin, Opt. Commun. \textbf{2}, 324 (1970).

\bibitem{4} J. J. Hopfield, Phys. Rev. \textbf{112}, 1555 (1958).

\bibitem{5} V. M. Agranovich and O. A. Dubovskii, JETP Lett. \textbf{3,} 223
(1966).

\bibitem{6} Ya. Aaviksoo, Ya. Lippmaa, and T. Reinot, Optics and
Spectroscopy (USSR) \textbf{62}, 419 (1987).

\bibitem{7} B. Deveaud, F. Clerot, N. Roy, K. Satzke, B. Sermage, D. S.
Katzer, Phys. Rev. Lett\textbf{\ 67}, 2355 (1991).

\bibitem{8} J. Knoester, Phys. Rev. Lett.\textbf{\ 68}, 654 (1992).

\bibitem{9} G. Bj\"{o}rk , S. Pau, J. M. Jacobson, H. Cao, and Y. Yamamoto,
Phys. Rev. \textbf{B 52}, 17310(1995).

\bibitem{10} D. S. Citrin, Phys. Rev.\textbf{B 47,} 3832 (1993).

\bibitem{11} V. M. Agranovich, D. M. Basko, and O. A. Dubovsky, J. Chem.
Phys. \textbf{106,} 3896 (1997).

\bibitem{12} A. Nakamura, H. Yamada, and T. Tookizaki, Phys. Rev. \textbf{B
40}, 8585 (1989).

\bibitem{13} E. Hanamura, Phys. Rev. \textbf{B 38}, 1228 (1988).

\bibitem{14} F. C. Spano, J. R. Kuklinski, and S. Mukamel, Phys. Rev. Lett. 
\textbf{65,} 211 (1990).

\bibitem{15} Y. C. Lee, D. S. Chuu, and W. N. Mei, Phys. Rev. Lett. \textbf{%
69,} 1081 (1992).

\bibitem{16} D. S. Citrin, Phys. Rev. Lett. \textbf{69}, 3393 (1992); D. S.
Citrin, Phys. Rev. \textbf{B 48}, 2535 (1993).

\bibitem{17} Y. N. Chen and D. S. Chuu, Phys. Rev. \textbf{B 61}, 10815
(2000); Europhys. Lett. \textbf{54}, 366 (2001).

\bibitem{18} Y. C. Lee and D. L. Lin, Phys. Rev. \textbf{A 6}, 388 (1972).

\bibitem{19} P. S. Lee and Y. C. Lee, Phys. Rev. \textbf{A 8, }1727 (1973).

\bibitem{20} V.V. Popova, T.V. Teperika, N.J.M. Horingb, and T.Yu. Bagaeva,
Solid State Commun. \textbf{127}, 589 (2003).

\bibitem{21} One should note that Z mode emission drops to zero at normal
direction ($k_{\parallel }=0$).
\end{thebibliography}
\end{document}